\documentclass[%
 aip,
 jmp,%
 amsmath,amssymb,
 reprint,%
]{revtex4-2}

\usepackage[normalem]{ulem}
\usepackage{graphicx}
\usepackage{dcolumn}
\usepackage{bm}
\usepackage{xspace}
\usepackage[version=4]{mhchem}
\usepackage{geometry}
\preprint{AIP/123-QED}
\DeclareUnicodeCharacter{2212}{-}
\begin{document}
\title{Nanoscale ferroelectric programming of van der Waals heterostructures}

\author{Dengyu Yang}
\affiliation{Department of Physics, Carnegie Mellon University, Pittsburgh, PA 15213}
\affiliation{Department of Physics and Astronomy, University of Pittsburgh, Pittsburgh, PA 15260}
\affiliation{Pittsburgh Quantum Institute, Pittsburgh, PA, 15260}
\author{Qingrui Cao}
\affiliation{Department of Physics, Carnegie Mellon University, Pittsburgh, PA 15213}
\affiliation{Pittsburgh Quantum Institute, Pittsburgh, PA, 15260}
\author{Erin Akyuz}
\affiliation{Department of Physics, Carnegie Mellon University, Pittsburgh, PA 15213}
\affiliation{Pittsburgh Quantum Institute, Pittsburgh, PA, 15260}
\author{John Hayden}
\affiliation{Department of Materials Science and Engineering, The Pennsylvania State University, University Park, PA 16802}
\author{Josh Nordlander}
\affiliation{Department of Materials Science and Engineering, The Pennsylvania State University, University Park, PA 16802}
\author{Muqing Yu}
\affiliation{Department of Physics and Astronomy, University of Pittsburgh, Pittsburgh, PA 15260}
\affiliation{Pittsburgh Quantum Institute, Pittsburgh, PA, 15260}
\author{Ranjani Ramachandran}
\affiliation{Department of Physics and Astronomy, University of Pittsburgh, Pittsburgh, PA 15260}
\affiliation{Pittsburgh Quantum Institute, Pittsburgh, PA, 15260}
\author{Patrick Irvin}
\affiliation{Department of Physics and Astronomy, University of Pittsburgh, Pittsburgh, PA 15260}
\affiliation{Pittsburgh Quantum Institute, Pittsburgh, PA, 15260}
\author{Jon-Paul Maria}
\affiliation{Department of Materials Science and Engineering, The Pennsylvania State University, University Park, PA 16802}
\author{Benjamin M. Hunt}
\email{bmhunt@andrew.cmu.edu}
\affiliation{Department of Physics, Carnegie Mellon University, Pittsburgh, PA 15213}%
\affiliation{Pittsburgh Quantum Institute, Pittsburgh, PA, 15260}
\author{Jeremy Levy}
\email{jlevy@pitt.edu}
\affiliation{Department of Physics and Astronomy, University of Pittsburgh, Pittsburgh, PA 15260}
\affiliation{Pittsburgh Quantum Institute, Pittsburgh, PA, 15260}

\date{\today}

\begin{abstract}

The ability to create superlattices in van der Waals (vdW) heterostructures via moir\'e interference heralded a new era in the science and technology of two-dimensional materials.  Through precise control of the twist angle, flat bands and strongly correlated phases have been engineered. The precise twisting of vdW layers is in some sense a bottom-up approach--a single parameter can dial in a wide range of periodic structures. Here, we describe a top-down approach to engineering nanoscale potentials in vdW layers using a buried programmable ferroelectric layer. Ultra-low-voltage electron beam lithography (ULV-EBL) is used to program ferroelectric domains in a ferroelectric $\mathrm{Al_{1-x}B_{x}N}$ thin film through a graphene/hexagonal boron nitride (hBN) heterostructure that is transferred on top. We demonstrate ferroelectric field effects by creating a lateral p-n junction, and demonstrate spatial resolution down to 35 nm, limited by the resolution of our scanned probe characterization methods. This innovative, resist-free patterning method is predicted to achieve 10 nm resolution and enable arbitrary programming of vdW layers, opening a pathway to create new phases that are inaccessible by moir\'e techniques. The ability to ``paint" different phases of matter on a single vdW ``canvas" provides a wealth of new electronic and photonic functionalities.
\end{abstract}


\maketitle

\section{\label{sec:level1}Introduction}

Van der Waals (vdW) heterostructures have gained significant attention due to their distinctive and diverse properties in low-dimensional materials \cite{Geim2013}. Through tuning of various external parameters, efforts to configure and understand vdW quantum materials encompass superconductivity \cite{RN4044, doi:10.1126/science.aav1910}, Mott-like insulator states \cite{RN4043, tang_simulation_2020}, ferromagnetism at integer \cite{doi:10.1126/science.aaw3780, doi:10.1126/science.aay5533, li_quantum_2021} and fractional \cite{Park2023, PhysRevX.13.031037, lu_fractional_2024} fillings of the moir\'e bands, interlayer moir\'e excitons \cite{tran_evidence_2019, alexeev_resonantly_2019, seyler_signatures_2019, jin_observation_2019}, and more. To perform an enhanced comprehension and exploration of these properties, as well as for the fabrication of diverse nanoscale quantum devices, the importance of electrostatic manipulation with efficiency and low disorder cannot be overstated. However, achieving such precise and localized control poses substantial challenges, underscoring a critical area for research and technique development. Moiré patterns represent a powerful method for manipulating long-wavelength periodic potentials, establishing themselves as a significant area of study \cite{Andrei2020, Kennes2021, Mak2022}. This technique's success in uncovering various novel condensed matter phases has solidified its position in the field. A question that now arises is how we can extend the capability to generate new phases that are inaccessible from moir\'e techniques. A top-down method that introduces programming degree of freedom that allows the ability to create a wider range of both periodic and aperiodic structures, and combining them in a single device is needed.

Most top-down approaches to electrostatic gating of vdW layers rely on electron-beam lithography (EBL) to create nanoscale patterns. This approach has been used to produce electrostatically gated sharp edge states \cite{Li2018PRL, Li2018}, sharp p-n junctions \cite{Chen2016}, and gating through patterned dielectric substrates \cite{Li2021, Forsythe2018, sun2023signature, PhysRevLett.130.196201}. A method utilizing a synergistic approach with electron beam and backgate demonstrates patterning of van der Waals materials with 200 nm resolution\cite{Shi2020}. In addition to EBL methods, atomic force microscopy (AFM) anodic oxidation \cite{Cohen2023} has been used for precise patterning with approximately 50 nm resolution, and open-faced vdW materials can be programmed using atomic force microscopy \cite{Li2019, Lipatov2019}. The scanned probe methods are limited in that the lithography step works well only with monolayer graphene. In general, it has been challenging to recreate periodic structures that form naturally in moir\'e superlattices.  

Ferroelectric materials are appealing to modern nanoelectronics due to their non-volatile and switchable electric polarization. Recent reports show that solid solutions in the AlN-ScN \cite{Noor-A-Alam2019, Jiang2021}, AlN-BN \cite{Zhu2021, Hayden2021, Zhu2022}, and ZnO-MgO \cite{Ferri2021} families support ferroelectricity with polarization values between 80 and 150 µC/cm$^2$, with excellent polarization retention, and stability, to thicknesses at or below 10 nm. These materials can be processed under conditions that are more chemically and thermally compatible with many mainstream semiconductor platforms \cite{Noor-A-Alam2019, Jiang2021, Zhu2021, Hayden2021, Ferri2021}.
\begin{figure}[!ht]
    \centering
    \includegraphics[width=1\textwidth]{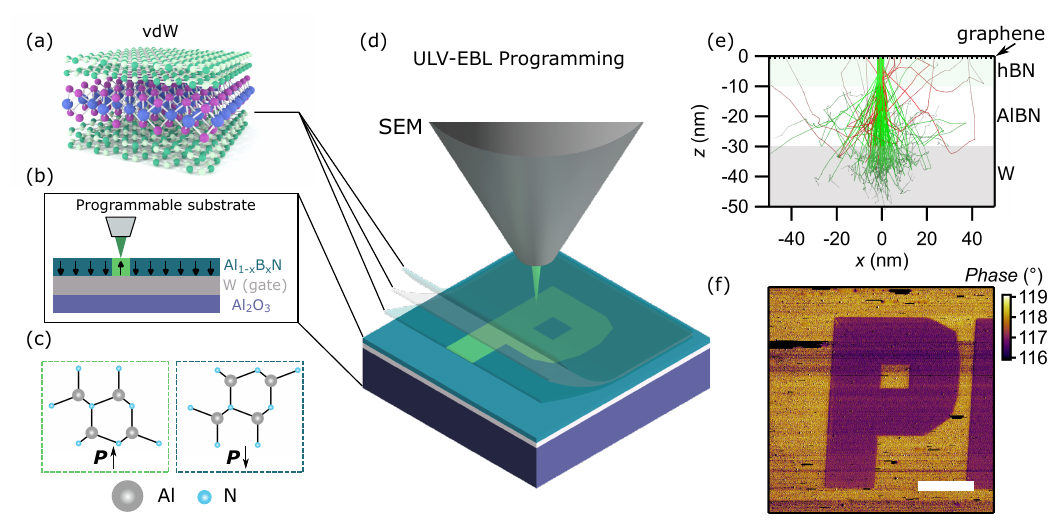}
    \caption[Ferroelectric switching using ULV-EBL]{\textbf{Ferroelectric switching using ULV-EBL} (a) Illustration of the vdW material atomic structures which is transferred onto the programmable substrate. (b) Illustration of ULV-EBL patterned ferroelectric polarization domains. (c) Atomic structure of the wurtzite Al(B)N of N-polar and Al-polar. (d) Schematic diagram of ferroelectric polarization switching with ULV-EBL. (e) Monte Carlo simulation for electron trajectories of acceleration voltage $V_\textrm{acc}=2$ kV at a graphene/hBN stack on 20 nm AlBN film. (f) AFM AC phase image of the ULV-EBL-exposed letter ``P". The scale bar represents 4 µm.}
    \label{fig1}
\end{figure}
As such, they present new possibilities for integration with 2D materials where programmed polarization patterns can tune electron transport.

Here, we describe an approach to programming ferroelectric thin films that are buried below a thick ($>30$ nm), multilayer van der Waals stack. The approach is based on a technique first developed by Nutt et al. \cite{Nutt1992} that utilized focused electron beams to switch ferroelectric polarization in LiNbO$_3$. We show that by carefully tuning the electron acceleration voltage, we can program the buried ferroelectric $\mathrm{Al_{1-x}B_{x}N}$ (AlBN) thin film with optimized spatial resolution particularly when using ultra-low voltage eletron beams \cite{Yang2020}. The acceleration voltage ($V_\textrm{acc}$) is optimized based on Monte Carlo simulations so that it is sufficient to penetrate through the ferroelectric film and minimize unwanted back-scattered electrons that can broaden the resolution. Our investigation also explores the impact of electron dose ($D$), and we successfully demonstrate beam conditions that produce feature sizes as small as 35 nm, dimensions which are comparable to the limit of our atomic force microscopy (AFM) AC scans and piezoelectric force microscopy (PFM) scans that measure them. We employ this method on the graphene vdW stack on AlBN and demonstrate graphene doping under selectively patterned polarization difference and show a p-n junction as a benchmark device. The method opens a pathway to integrate the ferroelectric film with other materials such as transition metal dichalcogenides (TMD). This new compatibility is aligned with the increasing complexity of current device concepts that can leverage intimate integration between ferroelectric material thin films, complex oxides, semiconductors and vdW materials to achieve multi-functional devices.

\section{Results}
 $\mathrm{Al_{1-x}B_{x}N}$ ($x$ =0.07) thin films with a target thickness of 11 nm or 20 nm are grown by dual-cathode reactive magnetron sputtering on W (40 nm) coated $c$-axis Al$_2$O$_3$ substrates at 300 °C. These films exhibit robust ferroelectric behavior with a switchable polarization as large as 130 µC/cm$^2$ which leads to a surface charge density close to 10$^{15}$ cm$^{-2}$ ~\cite{Zhu2021, Hayden2021, Zhu2022}.

As Figure \ref{fig1} shows, a letter ``P" is polarization patterned onto the ferroelectric AlBN film using the ULV-EBL exposure to switch the dipole moment from polarization down to polarization up (Fig.\ref{fig1}(b-d)). The sputtering conditions and surface preparations of AlBN growth produce films with a uniform polarization down dipole orientation and a compensating negative surface charge mechanism \cite{Zhu2021, Hayden2021, Zhu2022}. The exposure area dose is 4050 µC/cm$^2$. We use CASINO Monte Carlo simulation \cite{Hovington1997, Drouin1997, Hovington1997_2} to simulate the trajectories ((Fig.\ref{fig1}(e) and S1) and determine the optimal electron acceleration voltage used $V_\textrm{acc}$ to energize the electrons. To penetrate most of the thickness of the AlBN film, a $V_\textrm{acc}$ = 500 V (1 kV) is used to expose the 11 nm (20 nm) AlBN film. In both cases, the electron energy is sufficient enough to switch the surface polarization and be confirmed under AFM and PFM. Higher acceleration voltage is needed when there is a vdW stack on top of AlBN. As Figure \ref{fig1}(e) shows, a 2 kV $V_\textrm{acc}$ can penetrate through a graphene/10 nm-hBN stack on top and reach AlBN. The polarization difference on the ferroelectric sample with different surface charges is spatially characterized using the AFM AC mode or PFM. The AFM AC mode scan shows signal contrast from the surface charging difference (Fig. \ref{fig1}(f)). The PFM scan further shows the contrast between the two different polarized regions (Fig. S3). As shown by Calderon et al., the switching mechanism occurs through a sequential inhomogenous path of localized polyhedral distortions \cite{Calderon2023}. These give rise to a local nonpolar transition phase that mediates the global transition.
\begin{figure}[!ht]
    \centering
    \includegraphics[width=1\textwidth]{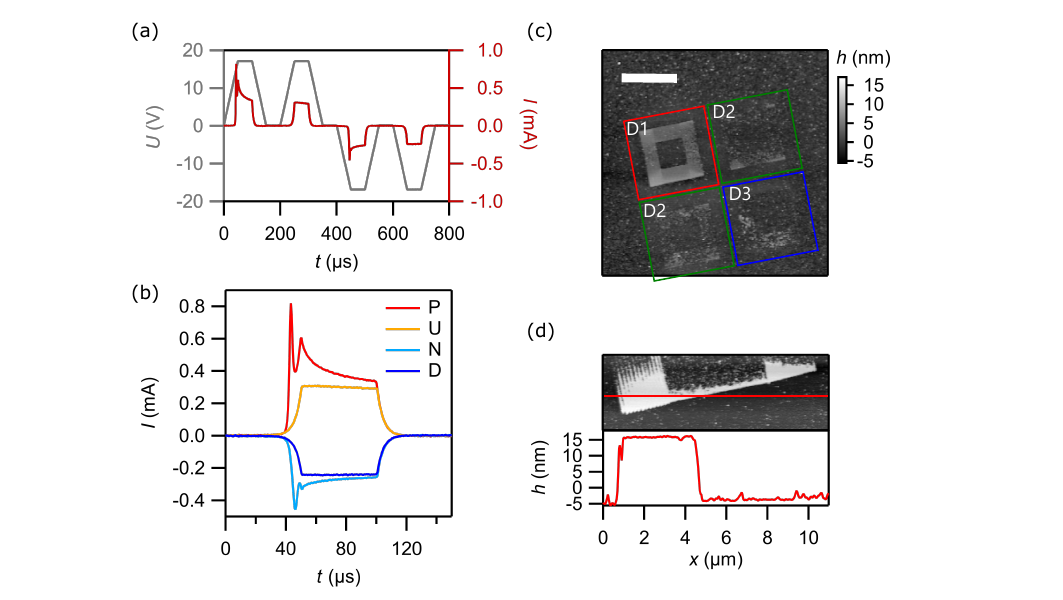}
    \caption[Characterization of AlBN ferroelectric and ULV-EBL patterned domains]{\textbf{Characterization of AlBN ferroelectric and ULV-EBL patterned domains.} (a) Positive up, negative down (PUND) measurements with applied voltage $U$ and measured current $I$ with respect to time. (b) Current measured under 17 V PUND voltage pulses. (c) AFM height scan of four square rings with different electron-beam doses, after a 30 s KOH etching described below. The red box is exposed with an area dose D1 = 51,200 µC/cm$^2$. The green boxes are exposed with an area dose D2 = 25,600 µC/cm$^2$. The blue box is exposed with an area dose D3 = 12,800 µC/cm$^2$. The scale bar denotes 10 µm. (d) AFM topography after etching with KOH for 60 seconds. The upper is the AFM height image for the partial red region in (c). The lower is a line cut along the red line in the upper image for its height profile. The measured etch depth is 20 nm which is the total AlBN thickness.}
    \label{fig2}
\end{figure}

While the AFM and PFM results strongly suggest polarization reversal, it is prudent to provide additional support given the companion artifacts that might occur using SPM-based measurements especially for such thin layers where leakage currents can be large. First we evaluate reference AlBN capacitors prepared from the identical parent films. Positive up negative down (PUND) measurements \cite{rabe_modern_2007} are widely regarded as a gold standard for separating ferroelectric polarization and leakage or dielectric contributions to switching (Fig.\ref{fig2}(a,b)). First we apply a +17 V voltage pulse to the bottom electrode of a 20 nm thick AlBN ferroelectric sample as Figure \ref{fig2}(a) shows. Since the sample polarization initially points down the first pulse (P) switches the polarization to up. An identical follow-up pulse (U) is then applied which should not switch the polarization. The measured current from U originates from only dielectric displacement and leakage currents. The identical test is carried for the negative side (N and D) except that the voltage sign is now the opposite. Figure \ref{fig2}(a) shows the current vs time traces for all PUND pulses. For the P and N pulses there are two features, a sharp spike associated primarily with polarization reversal and a decaying plateau associated with the polarization relaxation process and background leakage. We can draw this conclusion because the U and D pulses contain only the flat plateau feature. This testing ensures that the base material is indeed ferroelectric with a distinguishable polarization switching current.

We next leverage the etch selectivity between the two different polarizations to confirm that the polarization is switched by ULV-EBL. With the N-polar AlBN sample where the polarization is pointing down (Fig. \ref{fig1}(c)), the OH$^-$ react with Al$^+$ with the equation Eqs. (\ref{react1}, \ref{react2}). However, in Al-polar configuration, due to the negatively charged dangling nitrogen bond (Fig. \ref{fig1}(c)) which is supposed to be repulsive with OH$^-$, the etching is prevented \cite{ZHUANG20051, PINTO2022111753}. This gives rise to the different etching speed reacting to KOH:DI water liquid. We take advantage of this behavior to etch the sample with ULV-EBL electron-exposed and unexposed regions. To thoroughly electron-expose the 20 nm film, we use $V_\textrm{acc}$ = 2 kV based on the  the Monte Carlo simulation (Fig. \ref{fig1}e) which predicts electron penetration across the entire thickness. After that, an AZ400K (active component KOH 1:4 with water) is used to etch the AlBN film with the expectation that switched regions will be much more etch resistant than unswitched regions. The AFM scan shown in Figure \ref{fig2}(c) shows four features that were KOH etched for 30 s. The red box is exposed to an area dose of 51200 µC/cm$^2$. Green box ones are exposed with an area dose of 25600 µC/cm$^2$. The blue box is exposed to an area dose of 12800 µC/cm$^2$.  After this etch treatment only the highest dosed region retains the square box shape suggesting full switching. To further test this interpretation, the sample was KOH etched for another 30 seconds. Figure 2(d) shows a zoomed in AFM topography image and a linescan across highest electron-exposed feature  (Fig. \ref{fig2}(c, d)). The linescan shows a 20 nm step height, i.e., the entire film thickness, indicating complete polarization reversal (Fig. \ref{fig2}(d)). This is consistent with the other reports that C- surface has a much faster etching speed than the C+ surface of ferroelectric materials \cite{Nutt1992, ZHUANG20051, PINTO2022111753}, confirming  local and patternable electron beam induced switching.

\begin{figure}[h!]
    \centering
    \includegraphics[width=1\textwidth]{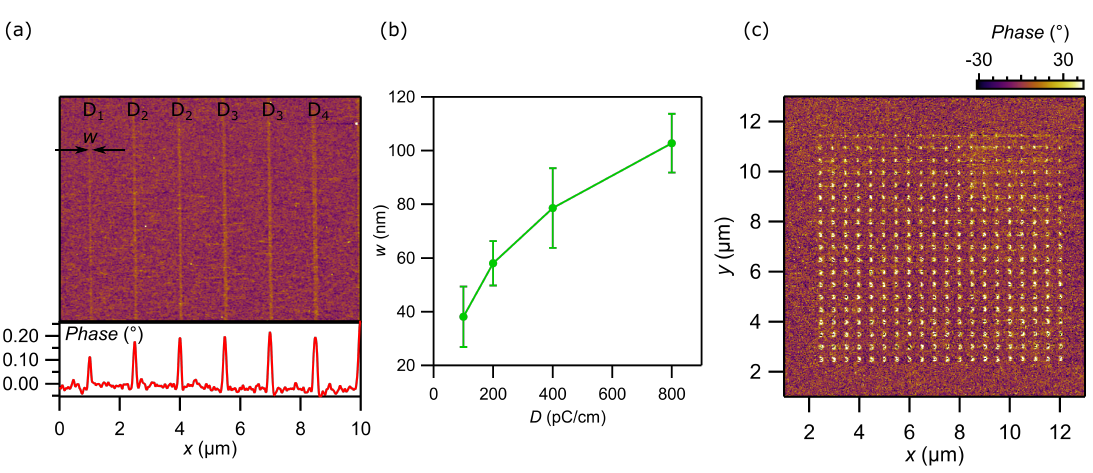}
    \caption[ULV-EBL ferroelectric switching resolution]{\textbf{ULV-EBL ferroelectric switching resolution.} (a) AFM phase image of a series of lines with different doses. Lines D$_1$, D$_2$, D$_3$, and D$_4$ are exposed with doses of 100, 200, 400, and 800 pC/cm, respectively. The scale bar is 2 µm. The lower image is a linecut to the upper scan. (b) Line width $w$ with respect to the exposure dose $D$. (c) PFM phase image of a square lattice with dose gradient from 0.01 pC (upper left) to 4 pC (bottom right) with 0.01 pC per step.}
    \label{fig3}
\end{figure}

We then explore the writing resolution and how the dose affects the measured width of line features, by exposing a series of lines at varying dose and characterizing them with AFM. Figure \ref{fig3} shows two different fine features, one-dimensional lines and a lattice made by dots. The lines in Figure \ref{fig3}(a) have different doses. Line D$_n$ has a dose of $D_\textrm{n}$ = 2$^{n-1} \times $100 pC/cm. Higher doses lead to wider exposure (Fig. \ref{fig3}(b)). This is understandable with the lateral distributions of electrons in interaction with the material. The smallest line width $w$ is 35 nm. Resolution is limited by both the e-beam resolution and possibly the smallest ferroelectric grain size. Figure \ref{fig3}(c) shows the PFM phase image of a square lattice with dose gradient from 0.01 pC (upper left) to 4 pC (bottom right) with 0.01 pC per step.

\begin{figure}[h!]
    \centering
    \includegraphics[width=1\textwidth]{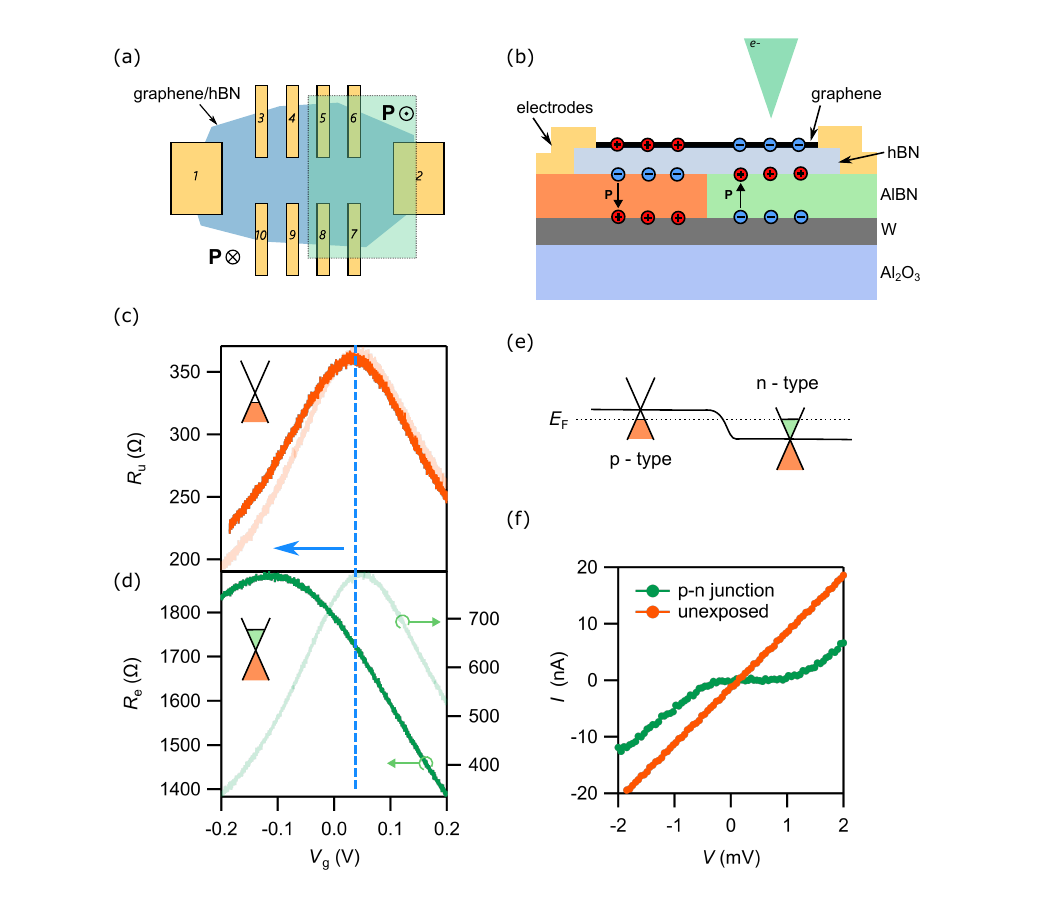}
    \caption[ULV-EBL ferroelectric switching on graphene]{\textbf{ULV-EBL ferroelectric switching on graphene} (a) Schematic diagram of the device. The blue region denotes the graphene/hBN device. There are electrical contacts to the graphene. The region covered in dashed line is the ULV-EBL exposure region. The rest of the Hall bar is the no exposure region. (b) Schematic diagram of the device geometry. A monolayer graphene and 10 nm thick hexagonal boron nitride van der Waals stack is on top of AlBN/W/sapphire substrate. (c)(d) $R$-$V_\textrm{g}$ measurement before (light color line) and after (opaque line) the exposure to show Dirac point for unexposed ($R_\textrm{u}$) and exposed ($R_\textrm{e}$) region. The blue dotted line denotes the original Dirac point position and the blue arrow shows the direction of the shift. $T$ = 300 K. (e) Illustration of the p-n junction energy band. Dashed line shows the Fermi energy $E_\textrm{F}$. (f) $I$-$V$ curves for the p-n junction and the unexposed region as comparison. $T$ = 15 mK.}
    \label{fig4}
\end{figure}

To integrate the AlBN ferroelectric with vdW materials, a monolayer graphene/hexagonal boron nitride (hBN) device (Fig. \ref{fig4}(a,b)) is fabricated onto the 20 nm AlBN substrate (See Methods). The hBN thickness is $\sim$ 10 nm for sample homogeneity. A 49-nm thick tungsten (W) layer underneath the AlBN serves as the gate of the ferroelectric field-effect transistor. For this device, an acceleration voltage $V_\textrm{acc} =$ 5 kV is chosen, based on Monte Carlo simulations, so that the electron beam penetrates through the vdW stack and switches the polarization of the ferroelectric thin film. The electron beam current is $I_\textrm{b} = $ 0.46 nA and the exposure dose $D = $ 50000 µC/cm$^2$. The graphene layer is kept grounded during the exposure. A solid region is exposed by ULV-EBL (dashed line in Fig. \ref{fig4}(a)). It is labeled the ``exposed" region (e) in comparison with the ``unexposed" region (u) where there is no ULV-EBL exposed on that half Hall bar region. In-situ graphene transport is performed, while the device is monitored in the ULV-EBL chamber. By applying the gate voltage $V_\textrm{g}$ within the coercive field range, graphene resistance $R$ is measured to identify the charge neutrality. Lead 6 is connected to lead 2, thus in the exposed side we process a three-terminal measurement (see supplementary materials for details).

Prior to exposure, graphene charge neutrality in $R_\textrm{e} \equiv V_\textrm{5-6}/I_\textrm{1-2}$ and $R_\textrm{u} \equiv V_\textrm{3-4}/I_\textrm{1-2}$ is both observed at $V_\textrm{g}^\textrm{CNP} = +0.08$ V, showing that graphene is initially hole doped (Fig. \ref{fig4}(c)). This observation is consistent with the fact that the ferroelectric thin film underneath the graphene is pre-polarized with a downward pointing polarization, implying the presence of negative surface charge that dopes graphene to p-type. Upon exposure, the exposed regions $R_\textrm{e}$ exhibit the charge neutrality at $V_\textrm{g}^\textrm{CNP} = -0.1$ V, while the charge neutrality position in the unexposed areas remains unchanged. The difference in $V_\textrm{g}^\textrm{CNP}$ in the exposed region is attributed to the alteration of the polarity in the regions irradiated by the electron beam, transitioning from a downward to an upward polarization. The resulting positive surface charge subsequently dopes graphene into n-type (Fig. \ref{fig4}(c,d)). Combined with the magneto-transport results, the corresponding change in carrier density upon ULV-EBL exposure is about $1.77 \times 10^{11}$ cm$^{-2}$ (see Supplemental Materials).

The patterning shown from Figure \ref{fig4}(c) and (d), where the device is half exposed and half unexposed, results in a p-n junction behavior in graphene when $V_\textrm{g}$ is applied such that the exposed region is n-doped and the unexposed region is p-doped. Figure \ref{fig4}(e) shows a schematic diagram of the p-n junction energy band in graphene. Using $I$-$V$ curve characterization, the p-n junction displays a diode-like barrier, in contrast to the linear current to voltage relation observed for the unexposed region (Fig. \ref{fig4}(f)). The p-n junction device serves as a fundamental component, demonstrating the potential to achieve more complex device geometries using this method.

\section{Discussions and Conclusions}

The approach developed here, involving a buried ferroelectric layer that is programmed using ULV-EBL to achieve electrostatic patterning of a vdW layer, offers many advantages. The programming step involves no cleanroom processing steps or e-beam resist, and is indefinitely stable after programming, unlike previous approaches involving LaAlO$_3$/SrTiO$_3$ where the programming is metastable~\cite{Huang2015, Yang2020}. Here, the effect is demonstrated in graphene, but it should be applicable to most 2D materials, e.g., transition-metal dichalcogenides (TMDs). The high spatial resolution afforded by ULV-EBL should enable periodic superlattices whose geometries are bounded only by lithographic lithographic limit thus providing new access to engineered bands and designer electronic phases. A future target is creating a platform for 2D analog quantum simulation, and thus predicting phases which have been discovered using other methods like moir\'e interference. Unlike the twistronic approach, device properties and materials could be mixed and matched from a single base material.

In the literature, there is a notable discrepancy wherein the experimentally measured carrier concentration in our device is orders of magnitude lower than the anticipated surface charge of a poled ferroelectric surface \cite{Hayden2021}. An immediate factor to consider is the impact of the device's architecture, which frequently presents an exposed graphene layer. In this geometry doping induced by the underlying polarization might be offset by surface physisorption, leading to a reduced carrier concentration. The same discrepancy is presently observed, and potentially originates from the exposed graphene.  Future experiments  will explore a fully hBN-encapsulated device where a top hBN can help prevent the charge compensation from other sources. Importantly, the thinness of the hBN cap will still allow direct polarization patterning by the scanning electron beam. Other concurrent passivation mechanisms may be present to reduce the net surface charge \cite{Baeumer2015}, such as in LiNbO$_3$, where charge compensation via surface reconstruction may be favored \cite{Levchenko2008}. In addition, the charge trapping effect \cite{doi:10.1021/nn201809k, park_charge_2022} is likely to manifest in graphene owing to the elevated surface charge density in AlBN and the thin hBN thickness.
    
In conclusion, we demonstrate the use of ULV-EBL to pattern the ferroelectric $\mathrm{Al_{1-x}B_{x}N}$ thin film where the electron beam energy and dose are optimized to generate domains as small as 35 nm, with a lower limit yet to be determined. The electron penetration depth is controlled precisely by $V_\textrm{acc}$ allowing one to regulate the electron dose in all three directions to minimize exposure of fragile substrates or other integrated layers that may experience deliterious effects. The electron-beam written polar pattern, and its compensating surface charge patter, can locally tune the charge density within an adjacent vdW layer with resolution on the order of 10s of nm. These observations and capabilities demonstrate the first and enabling step towards a new platform for a solid-state-based 2D analog quantum simulator.

\section{Methods}
\subsection{ULV-EBL}
In the experimental procedure, the Ultra-Low Voltage Electron Beam Lithography (ULV-EBL) was conducted using a Zeiss Gemini SEM 450 Scanning Electron Microscope (SEM), equipped with the high-speed 20~MHz Raith Elphy Plus lithography electronics system. The electron beam acceleration voltage has the range of 100 V to 30 kV. Unintended exposure was carefully avoided by making the adjustment and alignment of focus and stigmation on a standard Chessy chip. The ULV-EBL three-point alignment (TPA) and write-field alignment (WFA) were carried out at the specially designed markers using photolithography at the edge of the sample far away from the Hall bar region. Then, the beam was blanked and shuttled to the Hall bar to expose the designed pattern. The exposure was under vacuum with the chamber pressure $8.5 \times 10^{-7}$~mbar. The exposure working distance was 2.5 to 4 mm with the beam current varying from 57 pA to 460 pA. The dwell point step size was set to 10 nm.

\subsection{AFM and PFM}
The atomic force microscopy (AFM) and the piezoresponse force microscopy (PFM) were carried out under a commercial AFM system (Asylum Research MFP-3D). The AFM tip for AFM AC scan and contact scan was the doped silicon tip (Aspire Conical AFM Tips CFMR-25) with an in-air resonance frequency of 75 kHz and 3 N/m spring constant. The PFM was done with a PtIr$_5$-coated AFM probe from Nanosensors (PPP-EFM-50) with an in-air resonance of 75 kHz. The PFM image was done by driving the tip at a frequency in the range from 340 kHz to 370 kHz while engaging the sample surface and a driving voltage of 2 to 5 V. The scan was under ambient conditions.

\subsection{Device Fabrication}
\textit{AlBN/W/Al$_2$O$_3$ growth}: The commercial (001) Al$_2$O$_3$ substrates (Jiaozuo TreTrt Materials) was cleaned under isopropyl alcohol and methanol. Then a layer of 40 nm W (110) was grown on top while holding the substrate temperature at 300$^{\circ}$C using megnetron sputter dc sputtering under Ar condition. An AlBN film was deposited in the same chamber with reactive pulsed dc sputtering from an Al target and rf sputtering from a B target. \cite{Hayden2021, Zhu2021, Zhu2022}

\textit{Graphene/hBN device}: The graphene/hBN heterostructure was stacked by a modified dry transfer process based on polycaprolacetone (PCL) \cite{Son2020}, with a drop-down temperature at 80$^{\circ}$C. First, a monolayer graphene flake was exfoliated and identified on a $\mathrm{SiO_2/Si}$ substrate under optical microscope and picked up by a PCL stamp, followed by a thin hBN flake (thickness $\sim$ 10 nm) that served as an intermediate layer between the graphene and the AlBN substrate to improve sample homogeneity. Standard e-beam lithography was then used to define contacts to graphene; Cr/Au (2/100 nm) electrodes were deposited by e-beam evaporation to connect graphene and conductive pads to enable bonding to the sample. We employed silver-epoxy-based bonding to prevent punching through the 20 nm AlBN layer.

\subsection{KOH Etching}
We use photoresist developer AZ400K which active component is KOH 1:4 with water to etch the AlBN, simply by AZ400K bath at room temperature.
Under an alkaline condition for example KOH:DI water liquid here, there are reactions happening as:
\begin{align}
    \ce{2AlN + 3H2O &->[KOH] Al2O3 + 2NH3}\label{react1}\\
    \ce{AlN + 3H2O &->[KOH] Al(OH)3 + NH3}
    \label{react2}
\end{align}
where KOH is the catalyst shifting the equilibrium to the right.

\subsection{Electrical Measurements}
At room temperature, there is customized electrical feedthrough in ULV-EBL allowing \textit{in-situ} electrical measurement in the SEM chamber. The measurement channels consist of NI DAQ hardware (PXI-4461) and Krohn-Hite Preamplifiers (Model-7008). The Dirac point shift measurement is a lock-in measurement (frequency = 13 Hz) at room temperature. The sample was then transferred into a dilution refrigerator (Leiden MNK) and cooled down to 15 mK. A DC measurement for IV characterization of the p-n junction effect was carried out at the base temperature.

\begin{acknowledgments}
BMH, JL, J-PM and PRI acknowledge support from the Department of Energy under grant DOE-QIS (DE‐SC0022277).
Authors acknowledge Dmitry Shcherbakov for the helping with the drawing.
\end{acknowledgments}


\bibliography{Reference.bib}
\bibliographystyle{unsrt}

\end{document}


\maketitle
\tableofcontents
\newpage

\section{Monte Carlo Simulation}
\begin{figure}[h!]
        \centering
        \includegraphics[width=1\textwidth]{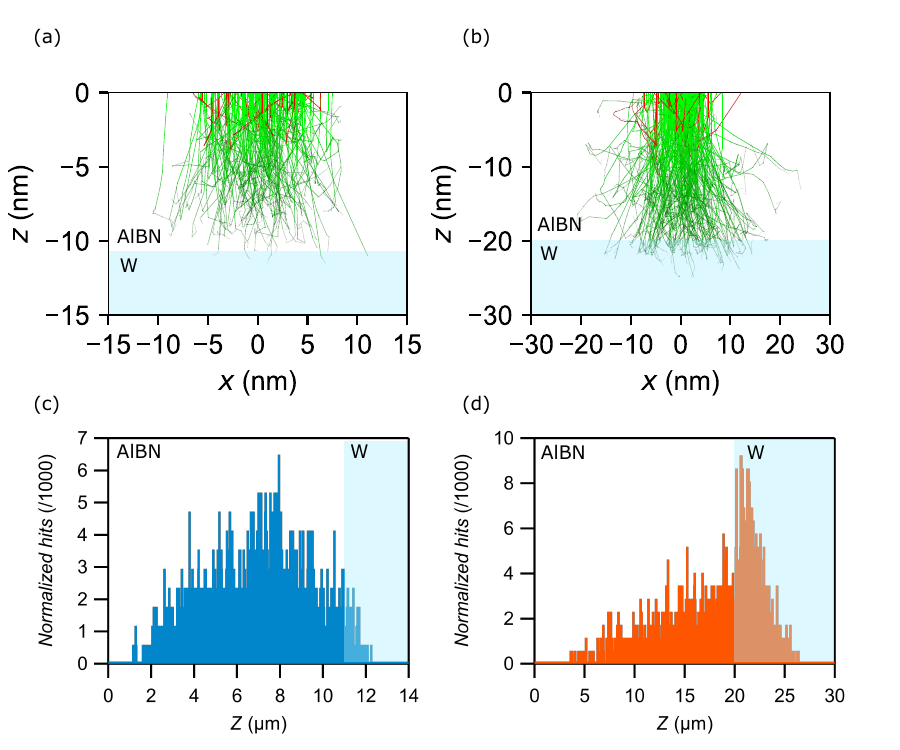}
        \caption[ULV-EBL ferroelectric switching penetration depth]{\textbf{Monte Carlo simulation of ULV-EBL ferroelectric switching penetration depth} (a) Monte Carlo simulation for electron trajectories of a 500 V acceleration voltage in an 11 nm AlBN film. (b) Monte Carlo simulation for electron trajectories of a 1 kV acceleration voltage in a 20 nm AlBN film. (c) Normalized electron stop-hit distribution with respect to $Z$ for a 500 V acceleration voltage at an 11 nm AlBN film. (d) Normalized electron stop-hit distribution with respect to depth $Z$ for a 1 kV acceleration voltage at a 20 nm AlBN film. The light blue regions denote the W layer underneath the AlBN.}
        \label{fig:supp1}
    \end{figure}

We use CASINO Monte Carlo simulation \cite{Hovington1997, Drouin1997, Hovington1997_2} to simulate the trajectories (Fig. \ref{fig:supp1} (a, b)) and determine the optimal electron acceleration voltage used $V_\textrm{acc}$ to energize the electrons. To penetrate most of the thickness of the AlBN film, a $V_\textrm{acc}$ = 500 V is used to expose the 11 nm AlBN film (Fig. \ref{fig:supp1} (a)). And a $V_\textrm{acc}$ = 1 kV is used to expose the 20 nm AlBN film (Fig. \ref{fig:supp1} (b)). In Figures \ref{fig:supp1} (a, b), the green lines show the trajectories of the penetrated electrons and the red lines show the backscattered electrons. Brighter colors denote higher energy and darker colors mean lower electron energy. In both cases, the electron energy is sufficient enough to switch the surface polarization and be seen under AFM. Figures \ref{fig:supp1} (c, d) show the normalized stop hits of the electrons with respect to depth $Z$. For the 11 nm AlBN film ($V_\textrm{acc}$ = 500 V) case, the electrons mostly stop before the AlBN/W interface. As in the case of 20 nm AlBN film ($V_\textrm{acc}$ = 1 kV), the electrons reach W but the electron energy decreases significantly due to the higher density of the material and the larger atomic number.

\newpage
\section{Landau fan measurements of graphene}

As a standard method to characterize the properties of the graphene, transport measurements of longitudinal resistance $R_\textrm{xx}$ and Hall resistance $R_\textrm{xy}$ in the quantum Hall regime are carried on the device and the Landau fan diagram is measured by applying the perpendicular magnetic field $B$ up to $\pm$18 T. The experiment is carried out at a dilution refrigerator, Leiden MNK, with a base temperature $<$ 15 mK. Due to the imperfect Hall geometry of the device, we measure the Landau fan diagram of both positive and negative magnetic field up to 18 T, and then carry out a symmetric analysis to both longitudinal resistance $R_\textrm{xx}$ and Hall resistance $R_\textrm{xy}$, where $R_\textrm{xx}(B) = 1/2(R_\textrm{xx} (B) + R_\textrm{xx} (-B))$ and $R_\textrm{xy}(B) = 1/2(R_\textrm{xy} (B) - R_\textrm{xy} (-B))$. 

\subsection{Unexposed region}

\begin{figure}[h!]
    \centering
    \includegraphics[width=1\textwidth]{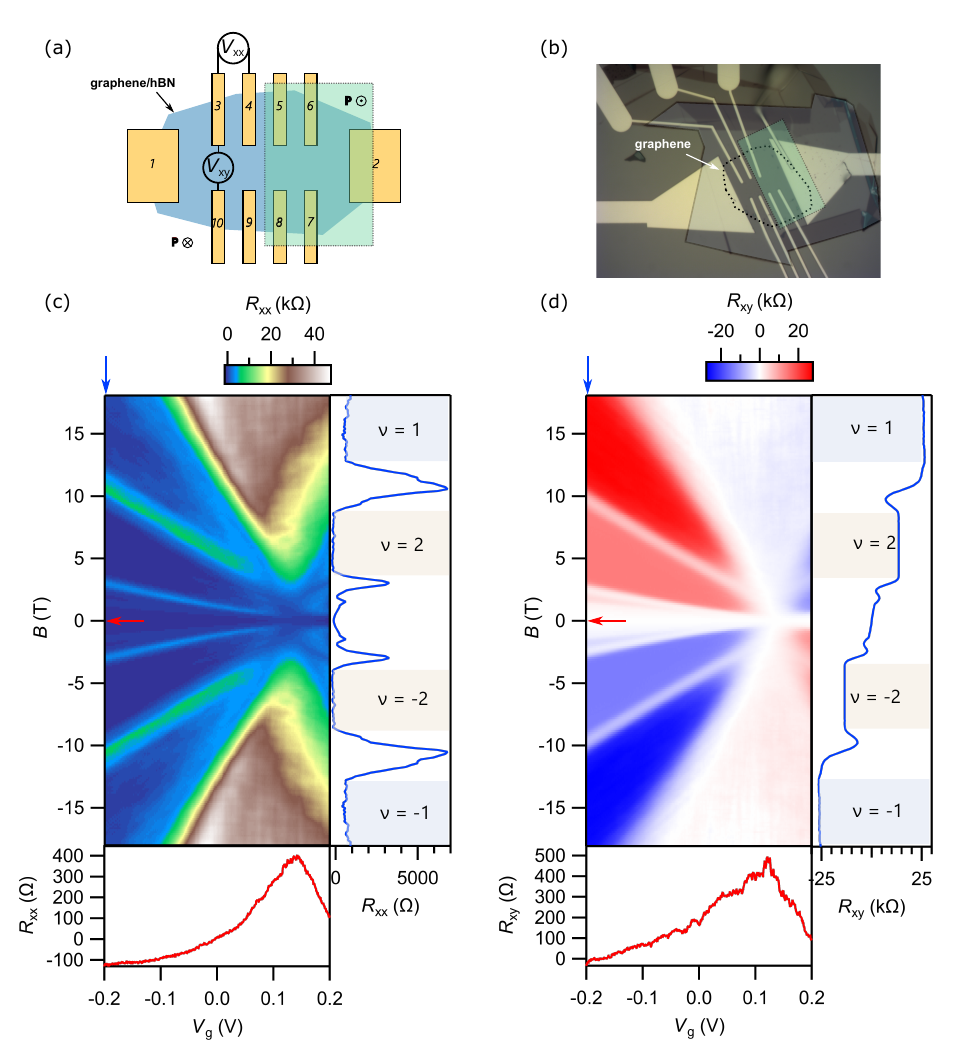}
    \caption[Graphene Landau fan diagram for unexposed region]{\textbf{Graphene Landau fan measurement for unexposed region} (a) Schematic diagram of the device. The blue region denotes the graphene/hBN device. There are electrical contacts to the graphene. The region covered in the dashed line is the ULV-EBL exposure region. The rest of the Hall bar is the unexposed region. The voltages of $R_\textrm{xx}$ and $R_\textrm{xy}$ are measured in electrode pairs of $V_\textrm{xx}$ and $V_\textrm{xy}$.} (b) Optical image of the device. Solid black like denotes the region of top hBN. The region covered in the dashed line under green color is the ULV-EBL exposure region. (c) Longitudinal resistance $R_\textrm{xx}$ as a function of the magnetic field $B$ and the gate voltage $V_\textrm{g}$ measured below 15 mK. (d) Hall resistance $R_\textrm{xy}$ as a function of the magnetic field $B$ and the gate voltage $V_\textrm{g}$ measured below 15 mK. In (c) and (d), right insets are a linecut taking at the blue arrow position ($V_\textrm{g} = $ -0.2 V) along magnetic field $B$, bottom insets are a linecut taking at the red arrow position ($B =$ 0) along the gate voltage $V_\textrm{g}$.
    \label{fig:supp2a}
\end{figure}

Figure \ref{fig:supp2a} shows the graphene Landau fan diagram of the unexposed region on the AlBN substrate. The longitudinal resistance $R_\textrm{xx} = V_\textrm{3-4}/I_\textrm{1-2}$ and the Hall resistance $R_\textrm{xy} = V_\textrm{3-10}/I_\textrm{1-2}$ are measured. Landau levels of $\nu = \pm 1, \pm 2$ can be observed. The Dirac point is observed at $V_\textrm{g}^\textrm{CNP} = $ +0.12 V. Compared with the data shown in the main text where the Dirac point is taken at 300 K, it shifts to the electron doping region a little. This is due to the device geometry which is an open-faced graphene device. When transferring from the ULV-EBL to the DR fridge, we need to break the SEM vacuum, transfer it in air and then load it into the DR fridge vacuum. The process that exposure to the air, introduces contact with the charges in the atmosphere so which shifts the Dirac point.

At $V_\textrm{g}= -0.2$ V (0.32 V away from the charge neutrality $V_\textrm{g}^\textrm{CNP}$), we observe the $\nu= \pm 2$ Landau gap at $B = \pm 6.5$ T, corresponding to a carrier density of $|n| = \frac{|\nu e B|}{h} \approx 3.14 \times 10^{11} \, \textrm{cm}^{-2}$. Based on this estimation, we deduce that the shift upon ULV-EBL writing ($|\Delta V_\textrm{g}| = 0.18$ V, see Fig. 4 in main text) is about $1.77 \times 10^{11} \, \textrm{cm}^{-2}$. Since the thicknesses of the $\mathrm{Al_{1-x}B_{x}N}$ substrate and hBN are 20 nm and 12 nm, respectively, with the hBN dielectric constant $\epsilon_{\mathrm{hBN}}=4$, we can further estimate that $\epsilon_{\mathrm{AlBN}} \approx 7.6$ using two parallel plate capacitors in series.

\subsection{Exposed region}

\begin{figure}[h!]
    \centering
    \includegraphics[width=1\textwidth]{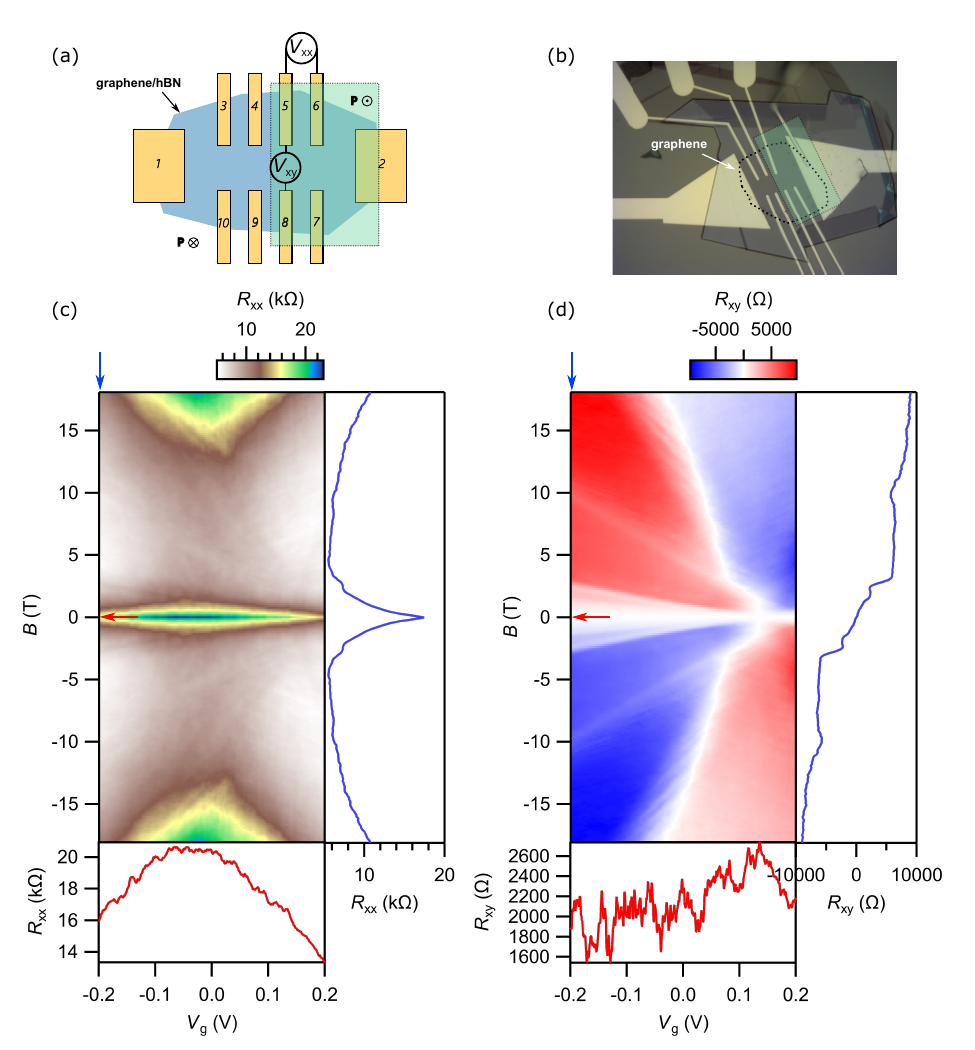}
    \caption[Graphene Landau fan diagram for exposed region]{\textbf{Graphene Landau fan measurement for exposed region} (a) Schematic diagram of the device. The blue region denotes the graphene/hBN device. There are electrical contacts to the graphene. The region covered in the dashed line is the ULV-EBL exposure region. The rest of the Hall bar is the unexposed region. The voltages of $R_\textrm{xx}$ and $R_\textrm{xy}$ are measured in electrode pairs of $V_\textrm{xx}$ and $V_\textrm{xy}$. (b) Optical image of the device. Solid black like denotes the region of top hBN. The region covered in the dashed line under green color is the ULV-EBL exposure region. (c) Longitudinal resistance $R_\textrm{xx}$ as a function of the magnetic field $B$ and the gate voltage $V_\textrm{g}$ measured below 15 mK. (d) Hall resistance $R_\textrm{xy}$ as a function of the magnetic field $B$ and the gate voltage $V_\textrm{g}$ measured below 15 mK. In (c) and (d), right insets are a linecut taking at the blue arrow position ($V_\textrm{g} = $ -0.2 V) along magnetic field $B$, bottom insets are a linecut taking at the red arrow position ($B =$ 0) along the gate voltage $V_\textrm{g}$.}
    \label{fig:supp2b}
\end{figure}

In the exposed region, $R_\textrm{xx} = V_\textrm{5-6}/I_\textrm{1-2}$ and $R_\textrm{xy} = V_\textrm{5-8}/I_\textrm{1-2}$ (Fig. \ref{fig:supp2b}). Lead \#7 is not working and lead \#6 and lead \#2 are connecting during fabrication process. It turns out the $R_\textrm{xx} = V_\textrm{5-6}/I_\textrm{1-2}$ here is a three-terminal measurement. Because of the switched ferroelectric polarization, Dirac point is now at $V_\textrm{g} = $ -50 mV showing in $R_\textrm{xx}$ at zero field (Fig. \ref{fig:supp2b} (c) bottom inset). But in $R_\textrm{xy}$ results, we cannot see a clear Dirac point position accordingly. That is because the $R_\textrm{xy}$ measured from lead \#5 and \#8 are too close to the separation edge between the exposed and unexposed region. Due to the imperfect Hall geometry of the device, it is hard to separate the exposed and unexposed results from $R_\textrm{xy}$. There is also a hint to this where the $R_\textrm{xy}$ Landau fan shows some different slopes that can be traced to a position different from the charge neutrality point (CNP)(Fig. \ref{fig:supp2b} (d)) indicating that the measured result is a mixing of two carrier density regions.
  
\section{More PFM scan images}

\begin{figure}[h!]
    \centering
    \includegraphics[width=1\textwidth]{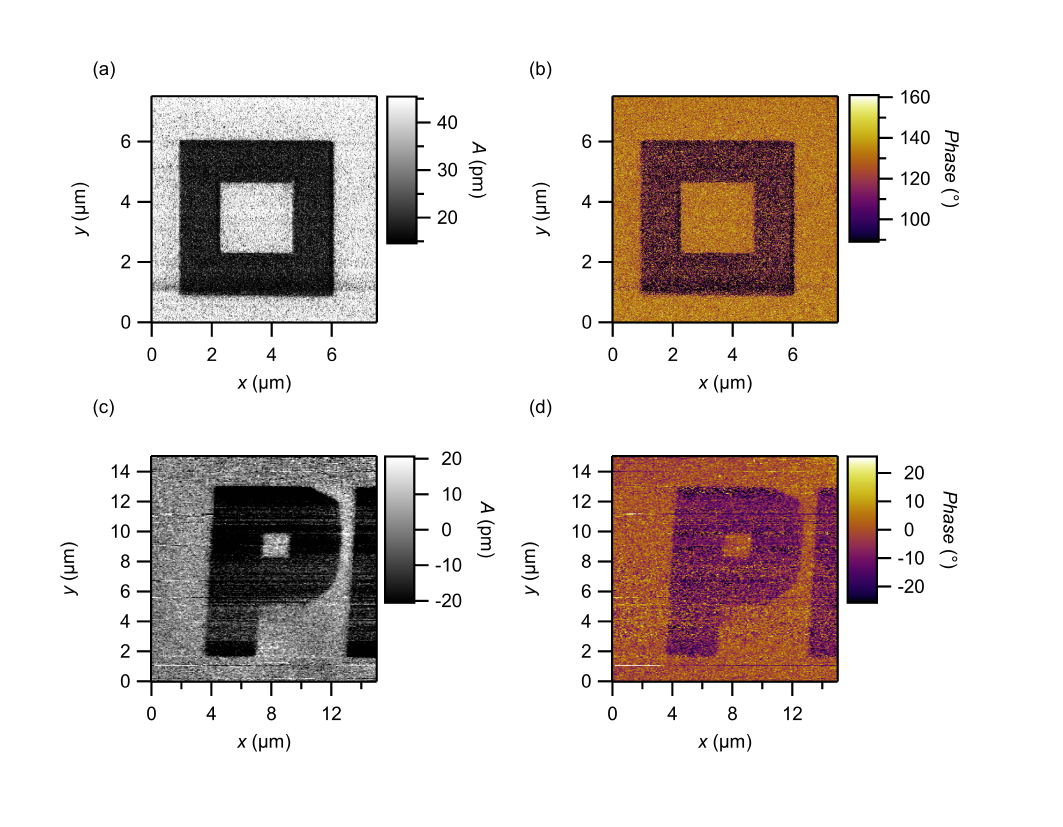}
    \caption[PFM images]{\textbf{PFM images} (a) PFM amplitude image. (b) PFM phase image. (c) PFM amplitude image. (d) PFM phase image.}
    \label{fig:supp3}
\end{figure}

We use piezoelectric force microscopy (PFM) to characterize the ULV-EBL patterned ferroelectric domains. As Figure \ref{fig:supp3} shows, a square annulus and the letter ``P" are exposed using ULV-EBL onto 20 nm thick AlBN thin film on top of W and sapphire substrate. Under PFM amplitude (Fig. \ref{fig:supp3} (a, c)) and phase (Fig. \ref{fig:supp3} (b, d)) images, the exposed annulus can be seen clearly. The amplitude gives a height change of near 30 pm coming from the piezoelectric response. The ULV-EBL switched ferroelectric domain shows a larger than 50 degree phase difference comparing with the AlBN intrinsic polarization inheriting from sample growth.

\bibliographystyle{unsrt}
\bibliography{Reference}